\documentclass[10pt,twocolumn,letterpaper]{article}

\usepackage[pagenumbers]{cvpr} 

\usepackage{iftex}
\ifXeTeX
  \usepackage{fontspec}
  \defaultfontfeatures{Ligatures=TeX}
\else
\ifLuaTeX
  \usepackage{fontspec}
  \defaultfontfeatures{Ligatures=TeX}
\fi\fi
\usepackage{booktabs}
\usepackage{multirow}

\usepackage{algorithm}
\usepackage{algpseudocode}
\definecolor{cvprblue}{rgb}{0.21,0.49,0.74}
\usepackage[pagebackref,breaklinks,colorlinks,allcolors=cvprblue]{hyperref}

\usepackage[textsize=tiny]{todonotes}
\newcommand{\myparagraph}[1]{{\vspace{.2em} \noindent \bf #1}}

\newcommand{\tcb}{\textcolor{blue}}

\newcommand{\tcr}{\textcolor{red}}

\def\paperID{*****} 
\def\confName{CVPR}
\def\confYear{2026}

\title{ZeroGVC: Zero-Shot Generative Video Compression with Autoregressive Diffusion Priors}

\author{
\textbf{Yixin Gao}\textsuperscript{1}\quad
\textbf{Xiaohan Pan}\textsuperscript{1}\quad
\textbf{Lin Liu}\textsuperscript{2}\thanks{Corresponding authors.}\quad
\textbf{Xin Li}\textsuperscript{1}\footnotemark[1]\quad
\textbf{Zhibo Chen}\textsuperscript{1}\footnotemark[1]\quad
\textbf{Qi Tian}\textsuperscript{2}\\
\textsuperscript{1}University of Science and Technology of China\quad
\textsuperscript{2}Huawei Inc.\quad
}

\begin{document}
\maketitle
\begin{abstract}
    Recent generative video compression methods leverage powerful generative priors to achieve perceptually pleasing reconstructions. However, most existing approaches require additional training to adapt generative models to produce realistic reconstructions from compact representations.
    In this paper, we propose ZeroGVC, a zero-shot generative video compression framework that leverages pretrained autoregressive diffusion priors for low-delay video reconstruction. ZeroGVC encodes the first frame of each group of pictures (GOP) with an image codec and represents subsequent P-frames through Codebook-Guided Autoregressive Latent Compression. This design is motivated by our observation that the compression scheme of denoising diffusion codebook models is effective in few-step consistency sampling. By selecting compact combinations of reproducible codebook noise vectors, ZeroGVC steers the latent denoising trajectory toward the target P-frame while allowing the decoder to reproduce the same trajectory in only a few denoising steps. In addition, we design an optional bidirectional reference mode that mitigates error propagation by leveraging the next I-frame context without introducing any additional bitrate overhead. Extensive experiments on standard video compression benchmarks demonstrate that ZeroGVC achieves superior perceptual reconstruction quality at ultra-low bitrates without any additional training.
\end{abstract}    
\section{Introduction}
\label{sec:intro}
Generative video compression (GVC) has recently emerged as a promising direction for producing realistic reconstructions at low bitrates, while previous neural video codecs often produce over-smoothed reconstructions~\cite{dcvc,dcvc-dc,dcvc-fm,dcvc-rt}. 
Instead of preserving all pixel-level details, GVC methods often leverage powerful generative priors to synthesize visually realistic content from compact representations. 
Some approaches introduce adversarial learning into neural video codecs to improve perceptual quality~\cite{mentzer2022neural,yangplvc}. More recently, some works explore more expressive representation spaces, such as discrete VQ-VAE tokens or continuous video latents, for ultra-low bitrate coding~\cite{qi2025generative_glcvideo,guo2025generative}. Some works exploit diffusion models to recover perceptually pleasing details from highly compact signals~\cite{ma2025diffusion,mao2025generative,zhang2026video,ma2026diffvc,ma2025diffvc,li2026yoda,xue2025single}. However, most existing methods require additional training to integrate such priors into video compression systems as reconstruction modules or auxiliary network components, incurring substantial computational cost and limiting their flexibility in adapting to increasingly powerful pretrained generative models.

Training-free generative compression offers an alternative route by directly reusing pretrained generative models without retraining. In the image compression field, several works formulate lossy compression by transmitting compact information associated with diffusion trajectories. DiffC~\cite{theis2022lossy} communicates noisy samples along the diffusion forward process through reverse-channel coding (RCC), and reconstructs images by denoising them with a pretrained diffusion model. Vonderfecht and Liu~\cite{vonderfecht2025lossy} further provide a more complete implementation of DiffC by resolving practical bottlenecks in reverse-channel coding. In parallel, DDCM~\cite{ohayon2025ddcm} and Turbo-DDCM~\cite{vaisman2026turboddcm} replace continuous Gaussian noise in the reverse diffusion process with selections from reproducible codebooks, and transmit only the compact indices to specify the denoising trajectory. Recent video extensions further demonstrate the potential of training-free GVC. Free-GVC~\cite{ling2026free} extends diffusion-based trajectory compression to video latents with temporal coherence based on RCC, while GVCC~\cite{zeng2026gvcc} converts the deterministic rectified-flow ODE of video diffusion models into an equivalent SDE, enabling DDCM-style codebook selection for zero-shot video compression. Despite this progress, a low-delay coding scheme for zero-shot generative video compression remains underexplored. Existing training-free methods operate at the group-of-pictures (GOP) level with multi-step denoising, which limits both decoding speed and first-frame latency, making them less compatible with low-delay video coding.

Recently, autoregressive (AR) video diffusion models have attracted growing attention for their ability to generate videos in a streaming manner~\cite{jin2024pyramidflow,chen2025skyreelsv2,yin2025causvid,huang2025selfforcing,zhu2026causalforcing}. These models combine causal temporal modeling with iterative denoising, rolling out frames sequentially conditioned on previously generated history. This causal, frame-by-frame generation paradigm is naturally aligned with low-delay video coding, as it enables sequential reconstruction using only previously decoded frames. Moreover, recent distillation and training strategies have enabled high-quality few-step AR generation~\cite{yin2025causvid,huang2025selfforcing,zhu2026causalforcing}. These observations motivate the question: can such pretrained AR diffusion models be leveraged for low-delay zero-shot generative video compression?

In this paper, we propose ZeroGVC, which encodes the first frame of each GOP using an image codec and reconstructs subsequent P-frames with a pretrained autoregressive diffusion model conditioned on previously decoded frames. The core of ZeroGVC is Codebook-Guided Autoregressive Latent Compression. Beyond the DDPM reverse chain considered by DDCM~\cite{ohayon2025ddcm} and Turbo-DDCM~\cite{vaisman2026turboddcm}, we find that the compression scheme of denoising diffusion codebook models is effective in few-step consistency sampling~\cite{song2023consistency}. Instead of transmitting dense residuals or latent features, ZeroGVC compresses the latent reconstruction trajectory itself. At each denoising step, the pretrained model first predicts a clean latent estimate from the current noisy state. The encoder then measures the target residual between this estimate and the target P-frame latent, and selects a sparse combination of atoms with quantized coefficients from a deterministic Gaussian codebook as the injected noise for the next step. Since the same codebook is regenerated at the decoder from a shared seed, only the selected atom-combination indices and quantized coefficients need to be transmitted. In addition, when the next I-frame is available, ZeroGVC can optionally use it as a bidirectional reference to improve reconstruction quality.

Our contributions are summarized as follows:
\begin{itemize}
    \item We introduce ZeroGVC, a training-free generative video compression framework that leverages pretrained autoregressive diffusion priors for low-delay video reconstruction. Unlike existing training-free generative codecs that operate at the GOP level, ZeroGVC reconstructs each P-frame sequentially from previously decoded frames.
     
    \item We find that the compression scheme of denoising diffusion codebook models is effective in few-step consistency sampling, and propose Codebook-Guided Autoregressive Latent Compression, which steers the autoregressive latent denoising trajectory toward the target P-frame by selecting and transmitting compact atom-combination indices and quantized coefficients. We further design an optional bidirectional reference mode to improve reconstruction quality when the next I-frame context is available.

    \item Extensive experiments validate the effectiveness of ZeroGVC across several standard video compression benchmarks, demonstrating strong perceptual reconstruction quality at ultra-low bitrates.
\end{itemize}

\section{Related Work}
\myparagraph{Generative Video Compression} \quad
Recent neural video compression (NVC) methods achieve impressive rate-distortion efficiency, but they often produce over-smoothed textures at low bitrates~\cite{dcvc,dcvc-dc,dcvc-fm,dcvc-rt}. To improve the perceptual quality of reconstructed videos, generative video compression (GVC) leverages powerful generative priors to produce visually pleasing reconstructions. Existing GVC methods can be broadly divided into training-based and training-free paradigms. 
Among training-based methods, GAN-based video codecs adopt adversarial learning~\cite{goodfellow2014generative} to produce more realistic reconstructions. Mentzer \etal~\cite{mentzer2022neural} first introduced GANs into neural video compression for detail synthesis and propagation, while Yang \etal~\cite{yangplvc} designed a recurrent conditional GAN with adversarial loss functions for perceptual learned video compression. More recently, some researchers have explored more suitable representation domains for ultra-low bitrate coding~\cite{qi2025generative_glcvideo,guo2025generative}. For example, GLC-Video~\cite{qi2025generative_glcvideo} compresses video latents into discrete VQ-VAE tokens, while Generative Latent Video Compression~\cite{guo2025generative} performs compression in the latent space of a continuous video tokenizer. Some works incorporate diffusion priors into video coding~\cite{ma2025diffusion,mao2025generative,zhang2026video}, using diffusion models to reconstruct perceptually pleasing videos from highly compact representations. To mitigate the heavy decoding cost of iterative denoising, several methods further adopt one-step diffusion reconstruction for faster perceptual video decoding~\cite{ma2025diffvc,xue2025single,li2026yoda}.

Training-free GVC has recently attracted attention because it can reuse pretrained video generative models without retraining. Building on RCC-based diffusion compression~\cite{theis2022lossy,vonderfecht2025lossy}, Free-GVC~\cite{ling2026free} reformulates extreme video compression as progressive coding of noisy latent representations in the diffusion space, with adaptive quality control and inter-GOP alignment to improve temporal coherence. GVCC~\cite{zeng2026gvcc} converts the ODE sampling process of rectified-flow video diffusion models into an equivalent SDE via the score-SDE framework~\cite{song2020score}, thereby enabling DDCM-style codebook selection~\cite{ohayon2025ddcm,vaisman2026turboddcm} for zero-shot video compression. However, existing training-free methods still face two practical issues: multi-step decoding limits reconstruction speed, and GOP-level generation is not naturally suited to streaming due to its first-frame latency. In contrast, ZeroGVC leverages pretrained few-step autoregressive diffusion priors for streaming P-frame compression, enabling high compression efficiency and low-latency zero-shot GVC.

\myparagraph{Autoregressive Video Diffusion Models} \quad
To enable streaming and low-latency video generation, recent works have explored autoregressive (AR) formulations for video diffusion, which naturally support progressive rollout over long temporal horizons. A representative direction combines AR temporal modeling with denoising diffusion~\cite{jin2024pyramidflow,chen2025skyreelsv2,deng2024nova,magi2025,yin2025causvid,huang2025selfforcing,zhu2026causalforcing}, where frames or chunks are generated sequentially along the temporal axis and each frame is synthesized through an iterative denoising process.
In these AR diffusion models, a key distinction lies in how the causal conditional distribution is trained: teacher forcing denoises each target frame conditioned on clean ground-truth history, while Diffusion Forcing (DF)~\cite{diffusionforcing} uses independently noised context and target tokens. To further improve generation efficiency, CausVid~\cite{yin2025causvid} adopts a DF-style training strategy and extends distribution matching distillation (DMD) to videos, distilling a pretrained bidirectional teacher into a few-step causal generator. However, both paradigms can still suffer from train-test mismatch, since training relies on ground-truth or noised context frames whereas inference conditions on previously generated outputs, leading to exposure bias and error accumulation.
Subsequently, Self Forcing~\cite{huang2025selfforcing} performs autoregressive self-rollout during training to better match the inference-time generation process. Causal Forcing~\cite{zhu2026causalforcing} further identifies an architectural gap in bidirectional-to-causal distillation and addresses it with AR-teacher-based ODE initialization before DMD.

\section{Preliminaries}
\label{sec:preliminaries}

\subsection{Denoising Diffusion Codebook Models}
\label{sec:preliminaries:ddcm}
Denoising Diffusion Codebook Models (DDCM)~\cite{ohayon2025ddcm} discretize the reverse process of Denoising Diffusion Probabilistic Models (DDPM)~\cite{ho2020denoising} by replacing its continuous Gaussian noise with entries from fixed, reproducible Gaussian codebooks. This turns a diffusion sampling trajectory into a sequence of discrete noise indices that can be stored or transmitted.

In DDPM, the forward process gradually corrupts a clean sample $\mathbf{x}_0$ through
\begin{equation}
    \mathbf{x}_{t} = \sqrt{\alpha_t}\mathbf{x}_{t-1}
    + \sqrt{1-\alpha_t}\boldsymbol{\epsilon}_t,
    \qquad \boldsymbol{\epsilon}_t \sim \mathcal{N}(\mathbf{0}, \mathbf{I}),
    \label{eq:ddpm_forward_step}
\end{equation}
and generation proceeds by reversing this discrete noising chain, which is commonly parameterized as
\begin{equation}
    \mathbf{x}_{t-1} = \mu_t(\mathbf{x}_t) + \sigma_t \boldsymbol{\eta}_t,
    \qquad \boldsymbol{\eta}_t \sim \mathcal{N}(\mathbf{0}, \mathbf{I}),
    \label{eq:ddpm_reverse_step}
\end{equation}
where $\mathbf{x}_t$ is the noisy latent at timestep $t$, $\mu_t(\cdot)$ is the learned reverse mean, and $\sigma_t$ is the noise scale. DDCM defines, for each timestep, a codebook
\begin{equation}
    \mathcal{C}_t = \left[\boldsymbol{\eta}_t^{(1)}, \boldsymbol{\eta}_t^{(2)}, \ldots, \boldsymbol{\eta}_t^{(K)}\right],
    \qquad \boldsymbol{\eta}_t^{(k)} \sim \mathcal{N}(\mathbf{0}, \mathbf{I}),
\end{equation}
which is sampled once and shared by the encoder and decoder. The reverse step then becomes
\begin{equation}
    \mathbf{x}_{t-1} = \mu_t(\mathbf{x}_t) + \sigma_t \mathcal{C}_t(k_t),
    \label{eq:ddcm_reverse_step}
\end{equation}
thereby yielding a finite yet expressive sampling space whose realization is specified by the selected index sequence $\{k_t\}$. Because the codebooks are fixed and reproducible, the decoder can reconstruct the same trajectory by deterministically re-selecting the transmitted indices.

For compression, the encoder selects codebook atoms that steer the generative trajectory toward a target signal $\mathbf{x}_0$. Let $\hat{\mathbf{x}}_{0|t}$ denote the clean estimate predicted from $\mathbf{x}_t$, and define the residual
\begin{equation}
    \mathbf{r}_t = \mathbf{x}_0 - \hat{\mathbf{x}}_{0|t}.
    \label{eq:target_residual}
\end{equation}
DDCM chooses the atom most aligned with this residual,
\begin{equation}
    k_t = \operatorname*{arg\,max}_{k \in \{1,\ldots,K\}}
    \left\langle \mathcal{C}_t(k), \mathbf{r}_t \right\rangle.
    \label{eq:ddcm_atom_selection}
\end{equation}
The resulting ordered index sequence constitutes the compressed representation of the target. DDCM further improves the representational capacity of each step through a matching-pursuit refinement, where multiple atoms are greedily selected and combined with quantized coefficients to form the injected noise at each timestep. While this refinement extends the method beyond the single-index regime, it also increases the encoding cost due to its iterative search.

Turbo-DDCM~\cite{vaisman2026turboddcm} improves the efficiency of multi-atom noise combination by replacing matching pursuit with a thresholding-based selection rule. It approximates the residual with a sparse signed combination of codebook atoms,
\begin{equation}
    \mathbf{s}_t^{*}
    =
    \operatorname*{arg\,min}_{\mathbf{s}_t}
    \left\lVert \mathcal{C}_t \mathbf{s}_t - \mathbf{r}_t \right\rVert_2^2,
    \quad
    \left\lVert \mathbf{s}_t \right\rVert_0 = M,
    \quad
    (\mathbf{s}_t)_i \in \mathcal{V}\cup\{0\},
    \label{eq:turboddcm_sparse_selection}
\end{equation}
where $M$ is the number of selected atoms and $\mathcal{V}$ is a finite coefficient set. 
Since random Gaussian atoms are nearly orthogonal in high-dimensional spaces, Turbo-DDCM selects the $M$ atoms whose correlations with $\mathbf{r}_t$ have the largest magnitudes. The synthesized noise is then
\begin{equation}
    \boldsymbol{\eta}_t^{*}
    =
    \frac{\mathcal{C}_t \mathbf{s}_t^{*}}
    {\operatorname{std}(\mathcal{C}_t \mathbf{s}_t^{*})},
    \qquad
    \mathbf{x}_{t-1} = \mu_t(\mathbf{x}_t) + \sigma_t \boldsymbol{\eta}_t^{*}.
    \label{eq:turboddcm_noise}
\end{equation}
Because the order of the selected atoms does not affect the synthesized noise, Turbo-DDCM encodes the selected atoms as an unordered set, together with their quantized coefficients, thereby avoiding redundant ordered representations in the bitstream.

\subsection{Few-Step Sampling in Consistency Models}
\label{sec:preliminaries:consistency}
Consistency models~\cite{song2023consistency} learn mappings that are self-consistent along the probability-flow ODE trajectories of diffusion models. Specifically, a consistency model parameterizes a function $f_\theta(\mathbf{x}_t,t)$ that maps states on the same trajectory to a common endpoint. For two states $\mathbf{x}_t$ and $\mathbf{x}_{t'}$ on the same trajectory, this self-consistency is expressed as
\begin{equation}
    f_\theta(\mathbf{x}_t,t) \approx f_\theta(\mathbf{x}_{t'},t') \approx \mathbf{x}_\epsilon.
    \label{eq:consistency_property}
\end{equation}
Here, $\mathbf{x}_\epsilon$ denotes the endpoint at the smallest noise level $\epsilon$, which is treated as the clean sample in practice; consistency models further impose the boundary condition $f_\theta(\mathbf{x}_{\epsilon},\epsilon)=\mathbf{x}_{\epsilon}$.

A single-step sampler draws an initial noisy state $\hat{\mathbf{x}}_{T}$ at the largest noise level $T$ and predicts the clean sample as $f_\theta(\hat{\mathbf{x}}_{T},T)$. For higher quality, consistency models also support few-step sampling by alternating clean prediction and noise reinjection over a decreasing sequence of noise levels $\tau_1>\tau_2>\cdots>\tau_{N-1}$, as summarized in Alg.~\ref{alg:cm_multistep}.
\begin{algorithm}[t]
\caption{Few-Step Consistency Sampling}
\label{alg:cm_multistep}
\begin{algorithmic}
    \Statex \textbf{Input:} consistency model $f_\theta(\cdot,\cdot)$, noise levels $\tau_1>\tau_2>\cdots>\tau_{N-1}$, initial noisy state $\hat{\mathbf{x}}_{T}$
    \State $\mathbf{x} \gets f_\theta(\hat{\mathbf{x}}_{T}, T)$ \Comment{clean prediction}
    \For{$n=1$ to $N-1$}
        \State sample $\boldsymbol{\eta} \sim \mathcal{N}(\mathbf{0},\mathbf{I})$
        \State $\hat{\mathbf{x}}_{\tau_n} \gets \mathbf{x}+\sqrt{\tau_n^2-\epsilon^2}\boldsymbol{\eta}$ \Comment{noise reinjection}
        \State $\mathbf{x} \gets f_\theta(\hat{\mathbf{x}}_{\tau_n}, \tau_n)$ \Comment{clean prediction}
    \EndFor
    \Statex \textbf{Output:} $\mathbf{x}$
\end{algorithmic}
\end{algorithm}
Here $\mathbf{x}$ denotes the current clean estimate, and $\hat{\mathbf{x}}_{\tau_n}$ denotes the noisy state at noise level $\tau_n$. 

\section{Method}
\label{sec:method}

\subsection{Pipeline}
\label{sec:method:pipeline}
\begin{figure*}[t]
    \centering
    \includegraphics[width=0.95\linewidth]{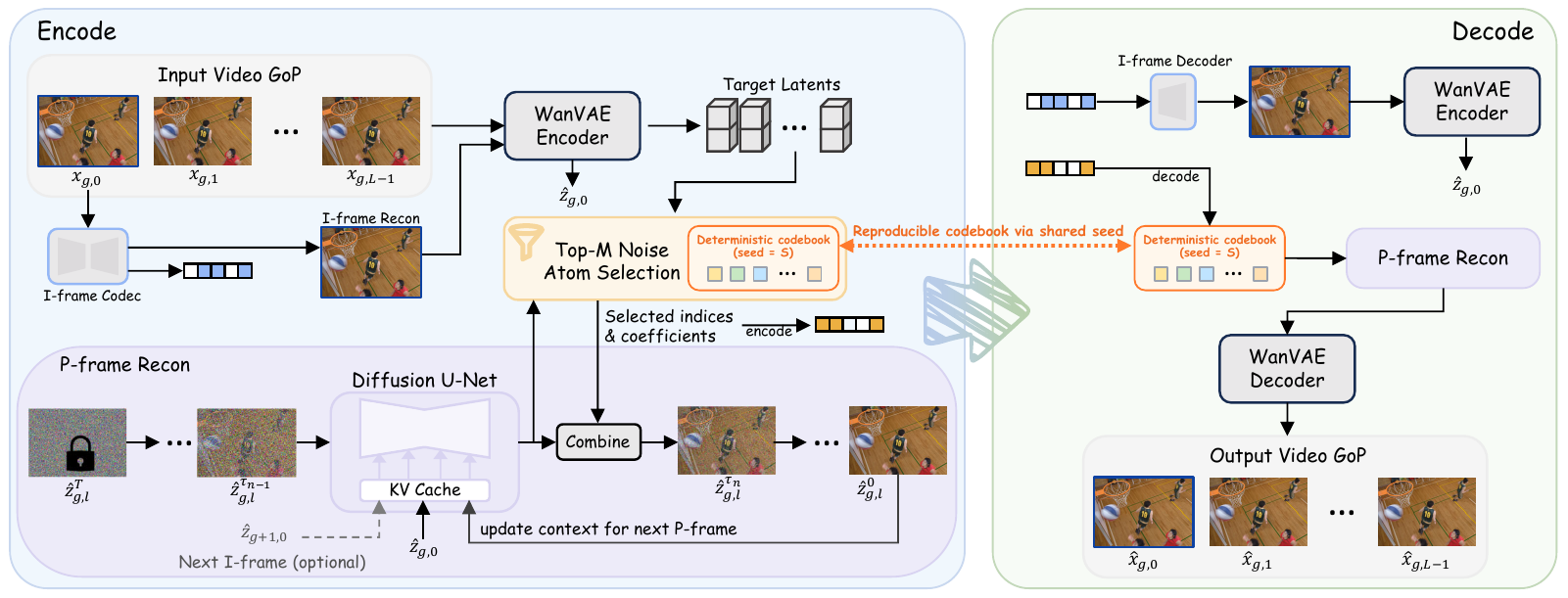}
    \vspace{-1mm}
    \caption{
        Overview of the proposed ZeroGVC framework. At the encoder, the first frame of each GOP is compressed by an image codec and mapped into the latent space to initialize the autoregressive context, while each subsequent P-frame is represented by compact atom-combination indices and quantized coefficients selected via Top-\(M\) Noise Atom Selection. At the decoder, the same codebook is regenerated from the shared seed, enabling reproducible latent reconstruction from the received atom-combination indices and quantized coefficients before WanVAE decoding. The optional bidirectional reference mode, shown by the gray dashed lines, uses the next decoded I-frame as auxiliary context to mitigate autoregressive error propagation without additional bitrate overhead.
    }
    \label{fig:framework}
\end{figure*}
We propose ZeroGVC, a training-free generative video compression framework that exploits a pretrained autoregressive diffusion prior for low-latency video reconstruction. As illustrated in Fig.~\ref{fig:framework}, given an input video, we divide it into GOPs and denote the $g$-th GOP as $\mathbf{X}_g=\{\mathbf{x}_{g,0},\mathbf{x}_{g,1},\ldots,\mathbf{x}_{g,L-1}\}$, where $\mathbf{x}_{g,0}$ is the I-frame and $\{\mathbf{x}_{g,l}\}_{l=1}^{L-1}$ are P-frames. The I-frame is encoded by a pretrained image codec and decoded as $\hat{\mathbf{x}}_{g,0}$. The WanVAE encoder then maps input video frames and the reconstructed I-frame into latent space, producing the reconstructed I-frame latent $\hat{\mathbf{z}}_{g,0}$ and target P-frame latents $\{\mathbf{z}_{g,l}\}_{l=1}^{L-1}$. The reconstructed I-frame latent is fed to the autoregressive diffusion model to initialize the KV cache used for subsequent P-frame reconstruction.

For each target P-frame latent $\mathbf{z}_{g,l}$, the encoder performs codebook-guided autoregressive latent compression conditioned on the current KV cache. At every diffusion step, the pretrained model first predicts a clean latent estimate from the current noisy latent state. Based on this estimate, a sparse combination of atoms with quantized coefficients is selected from a deterministic Gaussian codebook to steer the next state toward $\mathbf{z}_{g,l}$. The codebook is generated from a shared seed $S$, so only the selected atom-combination indices and quantized coefficients are written into the bitstream. Finally, we obtain the reconstructed latent $\hat{\mathbf{z}}_{g,l}^{0}$ and feed it to the autoregressive model to update the KV cache for the next P-frame.

On the decoder side, the I-frame bitstream is first decoded into the reconstructed image $\hat{\mathbf{x}}_{g,0}$, which is then mapped by the WanVAE encoder to $\hat{\mathbf{z}}_{g,0}$ to initialize the autoregressive context, consistent with the encoder side. For each P-frame, the decoder regenerates the deterministic codebook from the shared seed $S$, reads the transmitted atom-combination indices and quantized coefficients, synthesizes the same selected noise, and follows the corresponding diffusion denoising process to reconstruct $\hat{\mathbf{z}}_{g,l}^{0}$. The reconstructed latent is then fed to the autoregressive model to update the cache for the next P-frame. Finally, WanVAE decodes $\{\hat{\mathbf{z}}_{g,l}^{0}\}_{l=0}^{L-1}$ into the reconstructed GOP $\hat{\mathbf{X}}_g=\{\hat{\mathbf{x}}_{g,0},\ldots,\hat{\mathbf{x}}_{g,L-1}\}$.

\subsection{Codebook-Guided Autoregressive Latent Compression}
\label{sec:method:latent_compression}
P-frame compression is achieved by compressing the latent reconstruction trajectory rather than the latent itself. To keep the reconstruction aligned with the target frame, we introduce codebook-guided latent compression into the inference process of causal video diffusion models~\cite{zhu2026causalforcing,huang2025selfforcing}. Although the sampling process differs from the DDPM reverse chain considered by DDCM~\cite{ohayon2025ddcm} and Turbo-DDCM~\cite{vaisman2026turboddcm}, we find that selecting and transmitting the injected noise to minimize the target residual remains effective in few-step consistency sampling (Alg.~\ref{alg:cm_multistep}), extending denoising diffusion codebook models to few-step generative reconstruction.

Instead of drawing a Gaussian noise vector at each reinjection step, ZeroGVC selects the noise from a deterministic codebook so that the stochastic trajectory becomes compressible. For the $l$-th P-frame in GOP $g$, let $\mathbf{z}_{g,l}$ be the target clean latent and let $\mathcal{H}_{g,l}$ denote the current KV cache induced by previously reconstructed latents. Starting from an initial noisy latent $\hat{\mathbf{z}}_{g,l}^{T}$, the reconstruction trajectory follows the noise levels $\tau_1>\tau_2>\cdots>\tau_{N-1}$. At step $n$, the current clean estimate is denoted by $\tilde{\mathbf{z}}_{g,l}^{0|\tau_{n-1}}$, with the first step using the prediction from $\hat{\mathbf{z}}_{g,l}^{T}$. The encoder measures the target residual as
\begin{equation}
    \mathbf{r}_{g,l}^{n}
    =
    \mathbf{z}_{g,l}-\tilde{\mathbf{z}}_{g,l}^{0|\tau_{n-1}},
    \label{eq:ar_target_residual}
\end{equation}
which instantiates the residual definition in Eq.~\ref{eq:target_residual} for the target latent $\mathbf{z}_{g,l}$. Given the deterministic Gaussian codebook $\mathcal{C}_{\tau_n}$ generated from the shared seed $S$, the encoder selects a sparse coefficient vector $\mathbf{s}_{g,l}^{n,*}$ following the thresholding rule of Turbo-DDCM~\cite{vaisman2026turboddcm}.
The selected noise $\boldsymbol{\eta}_{g,l}^{n,*}$ is synthesized following Eq.~\ref{eq:turboddcm_noise}. The next noisy state is obtained by reinjecting this selected noise into the clean prediction,
\begin{equation}
    \hat{\mathbf{z}}_{g,l}^{\tau_n}
    =
    \tilde{\mathbf{z}}_{g,l}^{0|\tau_{n-1}}
    +
    \sqrt{\tau_n^2-\epsilon^2}\boldsymbol{\eta}_{g,l}^{n,*}.
    \label{eq:ar_noise_reinjection}
\end{equation}
The resulting clean prediction $f_\theta(\hat{\mathbf{z}}_{g,l}^{\tau_n},\tau_n;\mathcal{H}_{g,l})$ is used as the current estimate for the next step.
Only the selected atom-combination index and its associated quantized coefficients are transmitted at each step. Since both the initial noisy latent and $\mathcal{C}_{\tau_n}$ are generated from the shared seed, the decoder synthesizes the same $\boldsymbol{\eta}_{g,l}^{n,*}$ and performs P-frame latent reconstruction consistently with the encoder.

\subsection{Bidirectional Reference Mode}
\label{sec:method:reference}

The basic mode reconstructs each P-frame using only past decoded latents. We additionally provide a reference mode, indicated by the gray dashed lines in Fig.~\ref{fig:framework}, that can use the next decoded I-frame as an auxiliary context when it is available. Specifically, for the $g$-th GOP, both encoder and decoder use the next I-frame latent $\hat{\mathbf{z}}_{g+1,0}$ as an additional attention context when reconstructing P-frames in $\mathbf{X}_g$. This shared future reference serves as a stable long-range anchor that mitigates error propagation and improves reconstruction quality in the autoregressive decoding process.

The reference is implemented through additional KV cache entries. A direct append operation, however, would assign the next I-frame an incorrect temporal position: the reference keys and values would be treated as if they immediately followed the current P-frame, while their actual position is at the beginning of the next GOP. To preserve the relative temporal positions encoded by rotary position embeddings (RoPE), we store the future reference keys before rotation and re-apply RoPE using their true future frame indices. This ensures that attention scores between current queries and future reference keys reflect their correct temporal displacement. The optional bidirectional reference mode changes only the available attention context and introduces no additional bitrate overhead.

\section{Experiments}
\label{sec:experiments}

\subsection{Experimental Settings}
\myparagraph{Evaluation Datasets} \quad
We evaluate our method on several widely used video compression benchmarks, including JVET Class B, C, D, and E~\cite{flynnCommonTestConditions}, UVG~\cite{mercat2020uvg}, and MCL-JCV~\cite{wang2016mcl}.

\myparagraph{Baseline Methods} \quad
We compare our method with several state-of-the-art video compression approaches, spanning traditional hybrid codecs HEVC~\cite{HEVC} and VVC~\cite{VVC}, neural codecs DCVC-FM~\cite{dcvc-fm} and DCVC-RT~\cite{dcvc-rt}, and the recent generative compression model GLC-Video~\cite{qi2025generative_glcvideo}. 

\myparagraph{Evaluation Settings} \quad
Following previous works~\cite{dcvc-dc,dcvc-fm}, we evaluate the first 96 frames of each video sequence, with the intra period set to 32. We adopt the low-delay encoding configuration for all evaluated methods. 
To comprehensively assess perceptual quality, we report LPIPS~\cite{lpips}, DISTS~\cite{dists}, FID~\cite{fid}, and DOVER~\cite{wu2023dover}. Following previous works~\cite{mentzer2020high,msillm}, FID is computed on $256\times256$ image patches. Unlike the other reference-based metrics, DOVER is a no-reference subjective quality metric that evaluates videos from both technical and aesthetic perspectives.
In addition, we report MS-SSIM~\cite{msssim} to evaluate reconstruction fidelity in the supplementary material. Runtime is measured on an NVIDIA A800 GPU.

\myparagraph{Implementation Details} \quad
We adopt Causal Forcing (CF) \& Causal Forcing++ (CF++)~\cite{zhu2026causalforcing} as the autoregressive diffusion priors. 
We apply OneDC~\cite{xue2025one}, a state-of-the-art perceptual image codec, to transmit the first frame of each GOP. For our model, we set $K=16384$ and $N \in \{2,4,6\}$, and vary $M$ within the range of 64 to 384. 
For high-resolution inputs, we employ a tiling strategy~\cite{stablecodec,stablesr}, where each frame is encoded into the latent space and partitioned into overlapping 480p tiles. The denoising model and Top-\(M\) noise atom selection are then applied independently to each latent tile. The selected atom-combination indices and quantized coefficients from all tiles are written into the bitstream, while the reconstructed latent tiles are merged with a Gaussian weighting map to form the full-frame latent. During decoding, the stored atom-combination indices and quantized coefficients are read from the bitstream and applied to each latent tile under the same tile layout. The reconstructed tiles are then aggregated with a Gaussian weighting map and converted back to pixels using tiled VAE decoding.

\begin{figure*}[t]
    \centering
    \includegraphics[width=\linewidth]{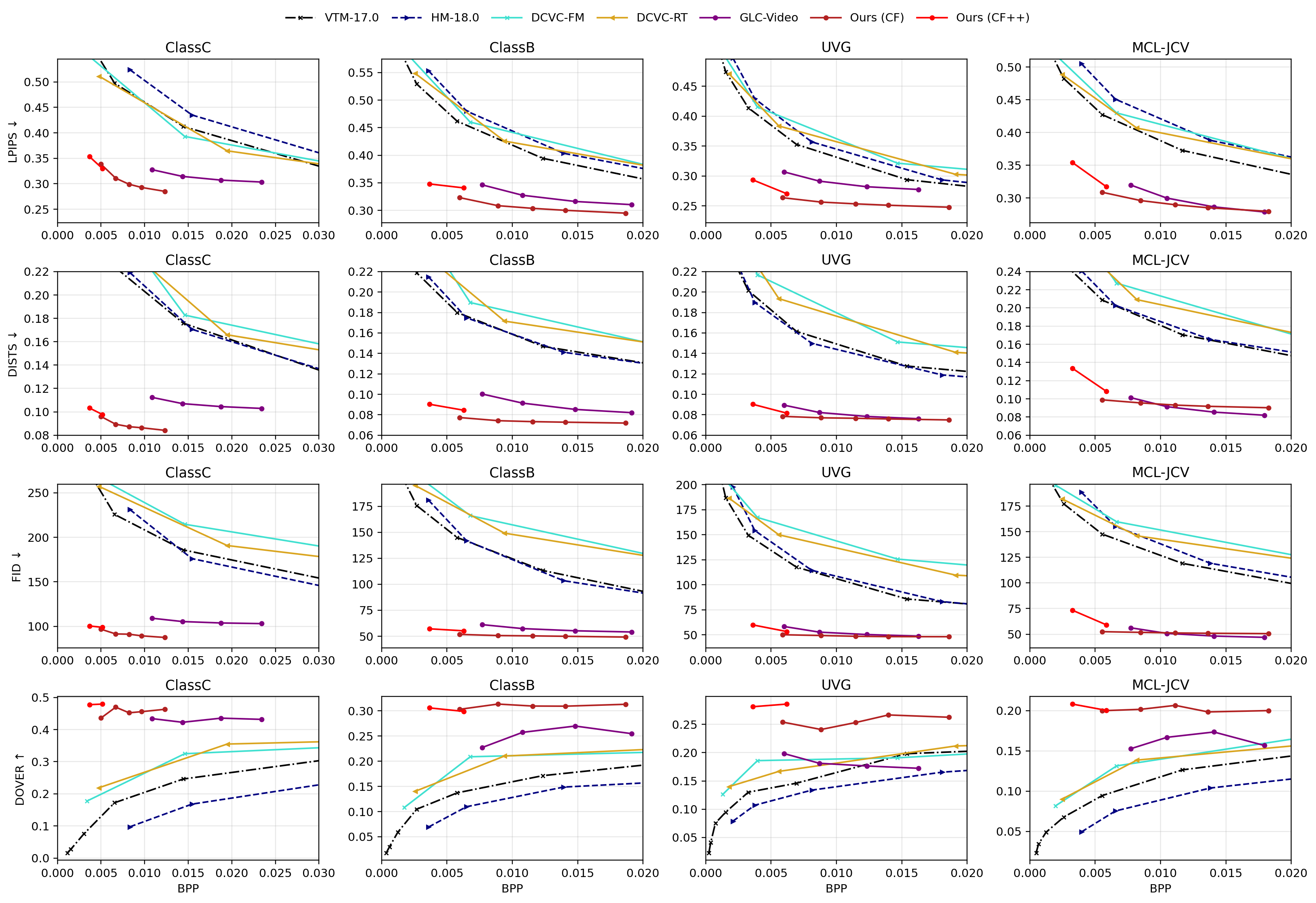}
    \vspace{-2em}
    \caption{Quantitative comparisons on HEVC Class C, HEVC Class B, UVG, and MCL-JCV.}
    \label{fig:main_results}
\end{figure*}
\subsection{Main Results}
\myparagraph{Quantitative Results} \quad
Fig.~\ref{fig:main_results} compares the perceptual performance on HEVC Class C, HEVC Class B, UVG, and MCL-JCV. On Class C, ZeroGVC achieves clear improvements over all competing methods across LPIPS, DISTS, FID, and DOVER. On the higher resolution datasets, including Class B, UVG, and MCL-JCV, ZeroGVC consistently achieves better LPIPS and DOVER scores than competing methods. It also maintains advantages in DISTS and FID on Class B and UVG. These results demonstrate the effectiveness of autoregressive diffusion priors for perceptual video compression.

We observe that the relative gain becomes narrower on high-resolution videos, especially at high bitrates. A possible reason is that inference on these datasets relies on tiled processing. Although tiling enables scalable reconstruction, processing local tiles independently may weaken the relative benefit of the generative prior. Nevertheless, ZeroGVC still performs favorably across most evaluation settings on Class B, UVG, and MCL-JCV, showing its effectiveness on both low-resolution and high-resolution videos.

\begin{figure*}[t]
    \centering
    \includegraphics[width=0.95\linewidth]{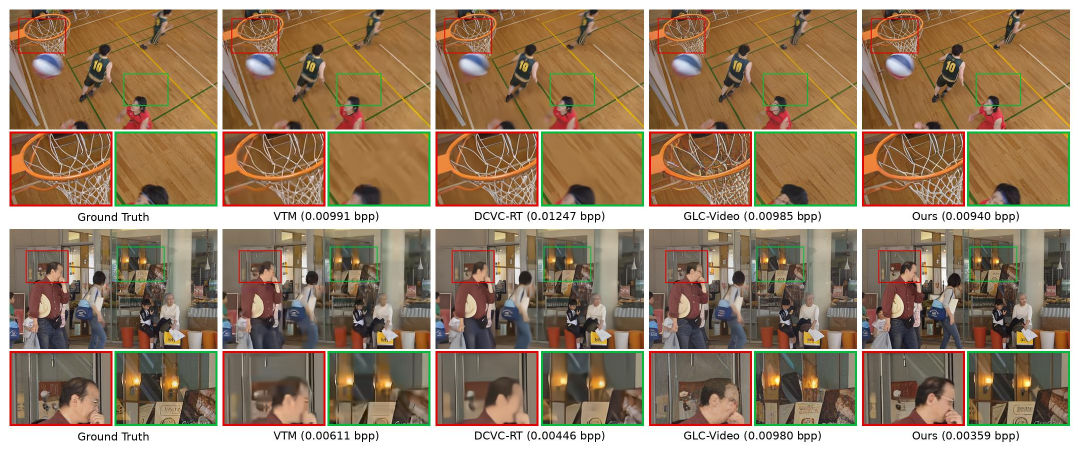}
    \vspace{-1em}
    \caption{Qualitative comparisons with baseline methods.}
    \vspace{-2mm}
    \label{fig:subjective_classc}
\end{figure*}

\myparagraph{Qualitative Results} \quad
Fig.~\ref{fig:subjective_classc} presents qualitative comparisons with baseline methods. At low bitrates, VTM and DCVC-RT produce over-smoothed reconstructions, where fine-grained structures and textures are noticeably degraded. GLC-Video improves perceptual realism in some regions but still exhibits visible artifacts and fails to faithfully reconstruct local details such as the basketball hoop structure and facial features. In contrast, even at comparable or lower bitrates, ZeroGVC preserves sharper structural details (\eg, the hoop rim and the athlete's posture in the top example, and the facial contours and indoor lamp in the bottom example) while better retaining background textures such as court markings and wall decorations. These results demonstrate that the autoregressive diffusion prior provides effective generative guidance for perceptual video reconstruction, enabling ZeroGVC to produce more realistic results at lower bitrates.
\begin{figure*}[t]
    \centering
    \includegraphics[width=0.85\linewidth]{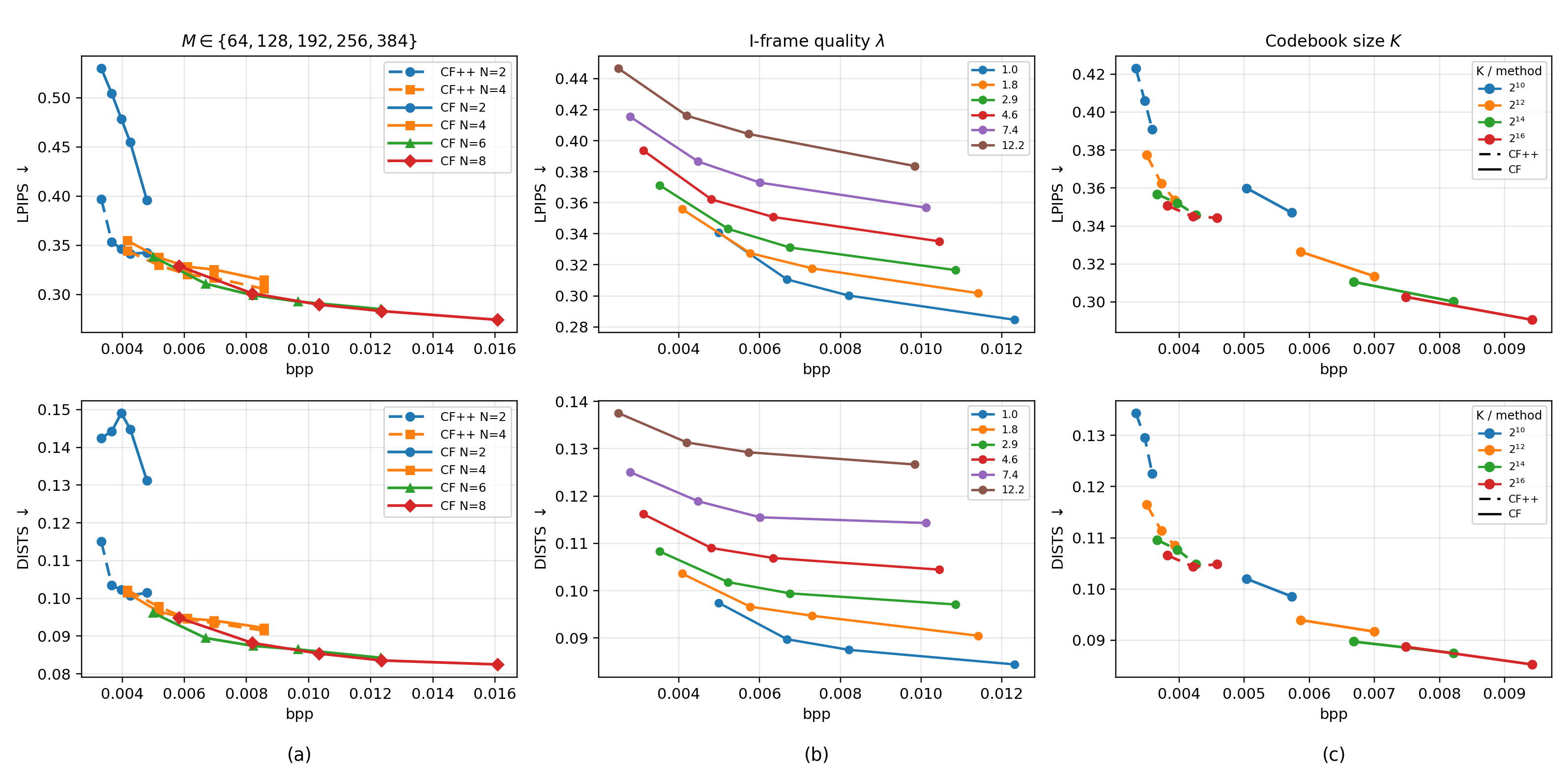}
    \vspace{-1em}
    \caption{Ablation studies on hyperparameters. We report LPIPS and DISTS rate-distortion curves under different (a) sampling steps $N$ and atom counts $M$, (b) I-frame quality levels $\lambda$, and (c) codebook sizes $K$.}
    \label{fig:ablation_classc_lpips_dists}
\end{figure*}

\myparagraph{Runtime} \quad
Table~\ref{tab:runtime} reports the encoding and decoding runtime on the HEVC Class C dataset. ZeroGVC achieves approximately 195/202 ms per frame for encoding/decoding at $N=6$, and the latency decreases to 96/124 ms per frame at $N=2$. Thanks to the few-step denoising design, ZeroGVC achieves efficient encoding and decoding, showing potential for practical usage while requiring no additional training.
\begin{table}[t]
    \centering
    \caption{Runtime of ZeroGVC under different autoregressive diffusion priors and denoising steps $N$. As runtime is largely insensitive to the atom count $M$ for a given prior and $N$, we report results averaged over the tested $M$ values.}
    \label{tab:runtime} 
    \small
    \begin{tabular}{@{}lccccc@{}}
        \toprule
        Prior & $N$ & Enc. (ms) & Enc. (FPS) & Dec. (ms) & Dec. (FPS) \\
        \midrule
        CF++ & 2 & 96.27  & 10.39 & 123.71 & 8.08 \\
        CF++ & 4 & 144.46 & 6.92  & 163.32 & 6.12 \\
        CF   & 6 & 195.04 & 5.13  & 201.74 & 4.96 \\
        \bottomrule
    \end{tabular}
    \vspace{-1.5em}
\end{table}

\subsection{Ablation Studies}
We conduct ablation studies on the HEVC Class C dataset.

\myparagraph{Sampling steps $N$ and atom count $M$} \quad
Fig.~\ref{fig:ablation_classc_lpips_dists}(a) studies the joint effect of the number of denoising steps $N$ and the selected atom count $M$. We sweep $N \in \{2,4,6,8\}$ and $M \in \{64,128,192,256,384\}$ under both CF++ and CF priors. Increasing $M$ enlarges the per-step correction capacity and improves perceptual quality at the cost of higher bitrate, while increasing $N$ offers more correction opportunities but introduces additional computational overhead. We select the configurations with the best overall perceptual quality as the operating points for final experiments. Specifically, for CF++, we use $(N,M)=(2,128),(4,128)$, while for CF we use $(N,M)=(6,64),(6,128),(6,192),(6,256),(6,384)$.

\begin{figure*}[t!]
    \centering
    \includegraphics[width=0.85\linewidth]{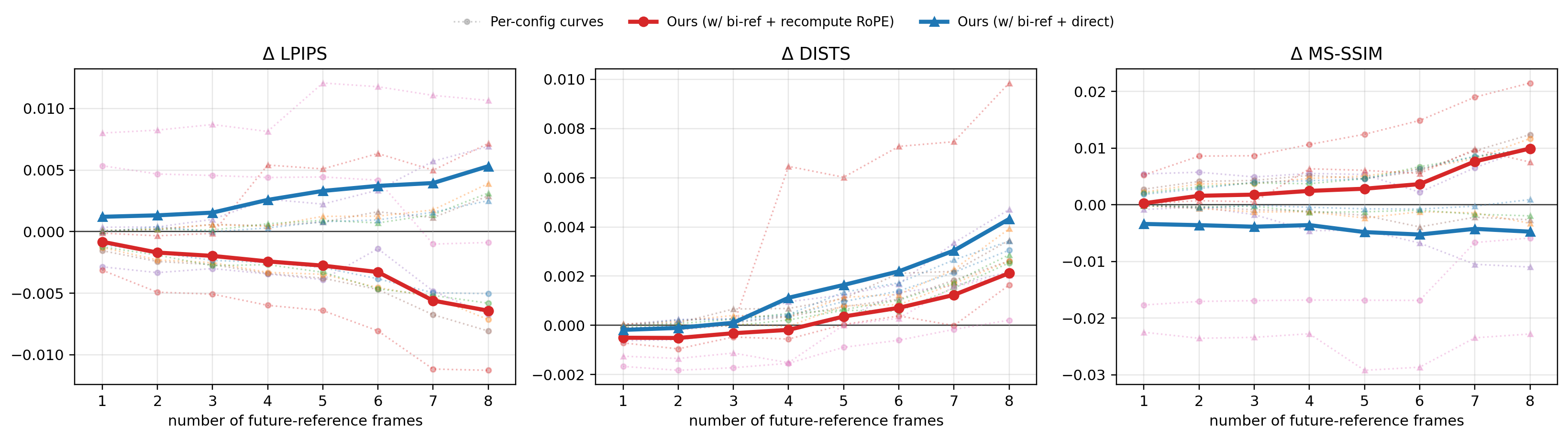}
    \vspace{-1em}
    \caption{Ablation study of the bidirectional reference mode under different numbers of future-reference latent frames. The dashed curves showing per-configuration results and solid curves showing the mean trend.}
    \label{fig:bidirectional_ablation}
    \vspace{-1em}
\end{figure*}

\begin{figure*}[t!]
    \centering
    \includegraphics[width=0.95\linewidth]{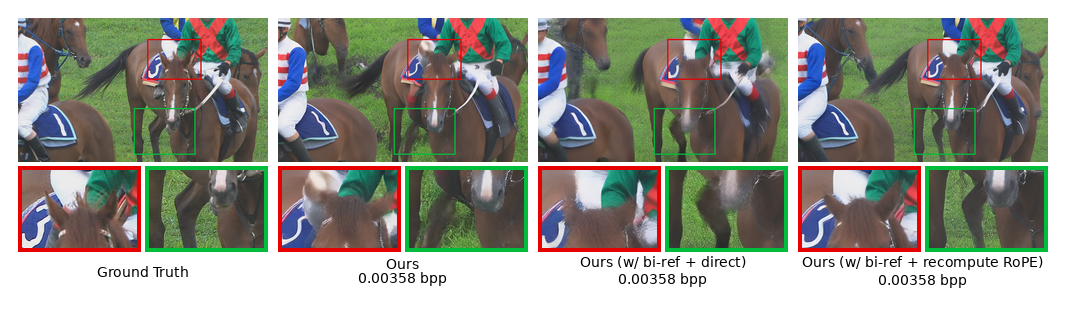}
    \vspace{-2em}
    \caption{Qualitative comparison of bidirectional reference mode. The future reference helps recover local structures and textures that are degraded by autoregressive error propagation.}
    \label{fig:bidirectional_subjective}
\end{figure*}
\myparagraph{I-frame Quality} \quad
We further ablate the quality level $\lambda$ of the image codec used for transmitting the first frame of each GOP. Since the decoded I-frame initializes the autoregressive context, its quality has a direct influence on all subsequent P-frame reconstructions. As shown in Fig.~\ref{fig:ablation_classc_lpips_dists}(b), a low-quality I-frame weakens the temporal anchor and causes error propagation across the GOP, leading to inferior LPIPS and DISTS even when more bits are spent on P-frame correction. In contrast, using $\lambda=1.0$ provides a reliable reference for the autoregressive prior and consistently yields the best performance. Therefore, we adopt $\lambda=1.0$ for all experiments.

\myparagraph{Codebook size $K$} \quad
The codebook size $K$ controls the expressiveness of the per-step noise search space. A small codebook restricts the available correction directions and leads to suboptimal reconstruction quality, while increasing $K$ improves the approximation of the target residual. However, the gain becomes marginal when the codebook is sufficiently large, and a larger $K$ also increases the inner-product search cost during encoding. Fig.~\ref{fig:ablation_classc_lpips_dists}(c) shows that $K=16384$ achieves a favorable balance between perceptual quality and computational cost. We therefore fix $K=16384$ in the main experiments.

\myparagraph{Bidirectional Reference Mode} \quad 
Fig.~\ref{fig:bidirectional_ablation} studies the effect of the number of reference latent frames in bidirectional reference mode. Compared with the baseline, incorporating the bidirectional reference brings clear overall gains, and recomputing RoPE for the reference tokens consistently outperforms direct KV-cache appending. As the number of reference frames increases, LPIPS improves continuously, while DISTS remains stable up to 4 frames but begins to degrade beyond that point. MS-SSIM exhibits consistent improvement across all reference counts. We therefore adopt 4 reference latent frames as the default setting, which provides a favorable trade-off between perceptual quality and reconstruction fidelity.

Fig.~\ref{fig:bidirectional_subjective} further provides qualitative evidence. With bidirectional references, the reconstructions better preserve structures such as the horse and rider boundaries. The results show that the future reference effectively alleviates autoregressive error propagation and improves reconstruction quality.


\section{Conclusion}
This paper proposes ZeroGVC, a training-free generative video compression framework that leverages pretrained autoregressive diffusion priors for low-delay video reconstruction. ZeroGVC encodes the first frame of each GOP with an image codec and reconstructs subsequent P-frames through Codebook-Guided Autoregressive Latent Compression, where compact atom-combination indices and quantized coefficients are transmitted to reproducibly steer the autoregressive latent denoising trajectory toward the target P-frame. We show that the compression scheme of denoising diffusion codebook models remains effective in few-step consistency sampling. In addition, we design an optional bidirectional reference mode that mitigates error propagation by leveraging the next I-frame context without introducing any additional bitrate overhead. Extensive experiments on standard benchmarks demonstrate that ZeroGVC achieves strong perceptual reconstruction quality at ultra-low bitrates without any additional training.
{
    \small
    \bibliographystyle{ieeenat_fullname}
    \bibliography{main}

\begin{thebibliography}{48}
\providecommand{\natexlab}[1]{#1}
\providecommand{\url}[1]{\texttt{#1}}
\expandafter\ifx\csname urlstyle\endcsname\relax
  \providecommand{\doi}[1]{doi: #1}\else
  \providecommand{\doi}{doi: \begingroup \urlstyle{rm}\Url}\fi

\bibitem[Bross et~al.(2021)Bross, Wang, Ye, Liu, Chen, Sullivan, and Ohm]{VVC}
Benjamin Bross, Ye-Kui Wang, Yan Ye, Shan Liu, Jianle Chen, Gary~J Sullivan, and Jens-Rainer Ohm.
\newblock Overview of the versatile video coding (vvc) standard and its applications.
\newblock \emph{TCSVT}, 2021.

\bibitem[Chen et~al.(2024)Chen, Mons{\'o}, Du, Simchowitz, Tedrake, and Sitzmann]{diffusionforcing}
Boyuan Chen, Diego~Mart{\'\i} Mons{\'o}, Yilun Du, Max Simchowitz, Russ Tedrake, and Vincent Sitzmann.
\newblock Diffusion forcing: Next-token prediction meets full-sequence diffusion.
\newblock In \emph{The Thirty-eighth Annual Conference on Neural Information Processing Systems}, 2024.

\bibitem[Chen et~al.(2025)Chen, Lin, Yang, Lin, Zhu, Fan, Zhang, Chen, Chen, Ma, et~al.]{chen2025skyreelsv2}
Guibin Chen, Dixuan Lin, Jiangping Yang, Chunze Lin, Junchen Zhu, Mingyuan Fan, Hao Zhang, Sheng Chen, Zheng Chen, Chengcheng Ma, et~al.
\newblock Skyreels-v2: Infinite-length film generative model.
\newblock \emph{arXiv preprint arXiv:2504.13074}, 2025.

\bibitem[Deng et~al.(2025)Deng, Pan, Diao, Luo, Cui, Lu, Shan, Qi, and Wang]{deng2024nova}
Haoge Deng, Ting Pan, Haiwen Diao, Zhengxiong Luo, Yufeng Cui, Huchuan Lu, Shiguang Shan, Yonggang Qi, and Xinlong Wang.
\newblock Autoregressive video generation without vector quantization.
\newblock In \emph{International Conference on Learning Representations}, pages 44730--44745, 2025.

\bibitem[Ding et~al.(2020)Ding, Ma, Wang, and Simoncelli]{dists}
Keyan Ding, Kede Ma, Shiqi Wang, and Eero~P Simoncelli.
\newblock Image quality assessment: Unifying structure and texture similarity.
\newblock \emph{IEEE transactions on pattern analysis and machine intelligence}, 44\penalty0 (5):\penalty0 2567--2581, 2020.

\bibitem[Flynn et~al.(16)Flynn, Sharman, and Rosewarne]{flynnCommonTestConditions}
D Flynn, K Sharman, and C Rosewarne.
\newblock Common test conditions and software reference configurations for hevc range extensions, document jctvc-n1006.
\newblock Joint Collaborative Team Video Coding ITU-T SG, 16.

\bibitem[Goodfellow et~al.(2014)Goodfellow, Pouget-Abadie, Mirza, Xu, Warde-Farley, Ozair, Courville, and Bengio]{goodfellow2014generative}
Ian Goodfellow, Jean Pouget-Abadie, Mehdi Mirza, Bing Xu, David Warde-Farley, Sherjil Ozair, Aaron Courville, and Yoshua Bengio.
\newblock Generative adversarial nets.
\newblock \emph{NeurIPS}, 27, 2014.

\bibitem[Guo et~al.(2025)Guo, Jia, Li, Zhang, Li, and Lu]{guo2025generative}
Zongyu Guo, Zhaoyang Jia, Jiahao Li, Xiaoyi Zhang, Bin Li, and Yan Lu.
\newblock Generative latent video compression.
\newblock \emph{arXiv preprint arXiv:2510.09987}, 2025.

\bibitem[Heusel et~al.(2017)Heusel, Ramsauer, Unterthiner, Nessler, and Hochreiter]{fid}
Martin Heusel, Hubert Ramsauer, Thomas Unterthiner, Bernhard Nessler, and Sepp Hochreiter.
\newblock Gans trained by a two time-scale update rule converge to a local nash equilibrium.
\newblock \emph{Advances in neural information processing systems}, 30, 2017.

\bibitem[Ho et~al.(2020)Ho, Jain, and Abbeel]{ho2020denoising}
Jonathan Ho, Ajay Jain, and Pieter Abbeel.
\newblock Denoising diffusion probabilistic models.
\newblock \emph{Advances in neural information processing systems}, 33:\penalty0 6840--6851, 2020.

\bibitem[Huang et~al.(2026)Huang, Li, He, Zhou, and Shechtman]{huang2025selfforcing}
Xun Huang, Zhengqi Li, Guande He, Mingyuan Zhou, and Eli Shechtman.
\newblock Self forcing: Bridging the train-test gap in autoregressive video diffusion.
\newblock \emph{Advances in Neural Information Processing Systems}, 38:\penalty0 167283--167308, 2026.

\bibitem[Jia et~al.(2025)Jia, Li, Li, Xie, Qi, Li, and Lu]{dcvc-rt}
Zhaoyang Jia, Bin Li, Jiahao Li, Wenxuan Xie, Linfeng Qi, Houqiang Li, and Yan Lu.
\newblock Towards practical real-time neural video compression.
\newblock In \emph{Proceedings of the Computer Vision and Pattern Recognition Conference}, pages 12543--12552, 2025.

\bibitem[Jin et~al.(2025)Jin, Sun, Li, Xu, Jiang, Zhuang, Huang, Song, Mu, and Lin]{jin2024pyramidflow}
Yang Jin, Zhicheng Sun, Ningyuan Li, Kun Xu, Hao Jiang, Nan Zhuang, Quzhe Huang, Yang Song, Yadong Mu, and Zhouchen Lin.
\newblock Pyramidal flow matching for efficient video generative modeling.
\newblock In \emph{International Conference on Learning Representations}, pages 23378--23402, 2025.

\bibitem[Li et~al.(2021)Li, Li, and Lu]{dcvc}
Jiahao Li, Bin Li, and Yan Lu.
\newblock Deep contextual video compression.
\newblock \emph{Advances in Neural Information Processing Systems}, 34:\penalty0 18114--18125, 2021.

\bibitem[Li et~al.(2023)Li, Li, and Lu]{dcvc-dc}
Jiahao Li, Bin Li, and Yan Lu.
\newblock Neural video compression with diverse contexts.
\newblock In \emph{Proceedings of the IEEE/CVF conference on computer vision and pattern recognition}, pages 22616--22626, 2023.

\bibitem[Li et~al.(2024)Li, Li, and Lu]{dcvc-fm}
Jiahao Li, Bin Li, and Yan Lu.
\newblock Neural video compression with feature modulation.
\newblock In \emph{Proceedings of the IEEE/CVF Conference on Computer Vision and Pattern Recognition}, pages 26099--26108, 2024.

\bibitem[Li et~al.(2026)Li, Zhang, Shi, Lu, and Ma]{li2026yoda}
Xingchen Li, Junzhe Zhang, Junqi Shi, Ming Lu, and Zhan Ma.
\newblock Yoda: Yet another one-step diffusion-based video compressor.
\newblock \emph{arXiv preprint arXiv:2601.01141}, 2026.

\bibitem[Ling et~al.(2026)Ling, Zhou, Li, Chen, Tian, Lu, and Zhang]{ling2026free}
Xiaoyue Ling, Chuqin Zhou, Chunyi Li, Yunuo Chen, Yuan Tian, Guo Lu, and Wenjun Zhang.
\newblock Free-gvc: Towards training-free extreme generative video compression with temporal coherence.
\newblock \emph{arXiv preprint arXiv:2602.09868}, 2026.

\bibitem[Ma and Chen(2025{\natexlab{a}})]{ma2025diffusion}
Wenzhuo Ma and Zhenzhong Chen.
\newblock Diffusion-based perceptual neural video compression with temporal diffusion information reuse.
\newblock \emph{ACM Transactions on Multimedia Computing, Communications and Applications}, 21\penalty0 (12):\penalty0 1--22, 2025{\natexlab{a}}.

\bibitem[Ma and Chen(2025{\natexlab{b}})]{ma2025diffvc}
Wenzhuo Ma and Zhenzhong Chen.
\newblock Diffvc-osd: One-step diffusion-based perceptual neural video compression framework.
\newblock \emph{arXiv preprint arXiv:2508.07682}, 2025{\natexlab{b}}.

\bibitem[Ma and Chen(2026)]{ma2026diffvc}
Wenzhuo Ma and Zhenzhong Chen.
\newblock Diffvc-rt: Towards practical real-time diffusion-based perceptual neural video compression.
\newblock \emph{arXiv preprint arXiv:2601.20564}, 2026.

\bibitem[Mao et~al.(2025)Mao, Cheng, Yang, Jin, and Ma]{mao2025generative}
Qi Mao, Hao Cheng, Tinghan Yang, Libiao Jin, and Siwei Ma.
\newblock Generative neural video compression via video diffusion prior.
\newblock \emph{arXiv preprint arXiv:2512.05016}, 2025.

\bibitem[Mentzer et~al.(2020)Mentzer, Toderici, Tschannen, and Agustsson]{mentzer2020high}
Fabian Mentzer, George Toderici, Michael Tschannen, and Eirikur Agustsson.
\newblock High-fidelity generative image compression.
\newblock \emph{arXiv preprint arXiv:2006.09965}, 2020.

\bibitem[Mentzer et~al.(2022)Mentzer, Agustsson, Ball{\'e}, Minnen, Johnston, and Toderici]{mentzer2022neural}
Fabian Mentzer, Eirikur Agustsson, Johannes Ball{\'e}, David Minnen, Nick Johnston, and George Toderici.
\newblock Neural video compression using gans for detail synthesis and propagation.
\newblock In \emph{European Conference on Computer Vision}, pages 562--578. Springer, 2022.

\bibitem[Mercat et~al.(2020)Mercat, Viitanen, and Vanne]{mercat2020uvg}
Alexandre Mercat, Marko Viitanen, and Jarno Vanne.
\newblock Uvg dataset: 50/120fps 4k sequences for video codec analysis and development.
\newblock In \emph{Proceedings of the 11th ACM multimedia systems conference}, pages 297--302, 2020.

\bibitem[Muckley et~al.(2023)Muckley, El-Nouby, Ullrich, J{\'e}gou, and Verbeek]{msillm}
Matthew~J Muckley, Alaaeldin El-Nouby, Karen Ullrich, Herv{\'e} J{\'e}gou, and Jakob Verbeek.
\newblock Improving statistical fidelity for neural image compression with implicit local likelihood models.
\newblock In \emph{International Conference on Machine Learning}, pages 25426--25443. PMLR, 2023.

\bibitem[Ohayon et~al.(2025)Ohayon, Manor, Michaeli, and Elad]{ohayon2025ddcm}
Guy Ohayon, Hila Manor, Tomer Michaeli, and Michael Elad.
\newblock Compressed image generation with denoising diffusion codebook models.
\newblock In \emph{Forty-second International Conference on Machine Learning}, 2025.

\bibitem[Qi et~al.(2025)Qi, Jia, Li, Li, Li, and Lu]{qi2025generative_glcvideo}
Linfeng Qi, Zhaoyang Jia, Jiahao Li, Bin Li, Houqiang Li, and Yan Lu.
\newblock Generative latent coding for ultra-low bitrate image and video compression.
\newblock \emph{IEEE Transactions on Circuits and Systems for Video Technology}, 2025.

\bibitem[Song et~al.(2020)Song, Sohl-Dickstein, Kingma, Kumar, Ermon, and Poole]{song2020score}
Yang Song, Jascha Sohl-Dickstein, Diederik~P Kingma, Abhishek Kumar, Stefano Ermon, and Ben Poole.
\newblock Score-based generative modeling through stochastic differential equations.
\newblock \emph{arXiv preprint arXiv:2011.13456}, 2020.

\bibitem[Song et~al.(2023)Song, Dhariwal, Chen, and Sutskever]{song2023consistency}
Yang Song, Prafulla Dhariwal, Mark Chen, and Ilya Sutskever.
\newblock Consistency models.
\newblock In \emph{Proceedings of the 40th International Conference on Machine Learning}, pages 32211--32252. PMLR, 2023.

\bibitem[Sullivan et~al.(2012)Sullivan, Ohm, Han, and Wiegand]{HEVC}
Gary~J Sullivan, Jens-Rainer Ohm, Woo-Jin Han, and Thomas Wiegand.
\newblock Overview of the high efficiency video coding (hevc) standard.
\newblock \emph{TCSVT}, 22\penalty0 (12):\penalty0 1649--1668, 2012.

\bibitem[Teng et~al.(2025)Teng, Jia, Sun, Li, Li, Tang, Han, Zhang, Zhang, Luo, et~al.]{magi2025}
Hansi Teng, Hongyu Jia, Lei Sun, Lingzhi Li, Maolin Li, Mingqiu Tang, Shuai Han, Tianning Zhang, WQ Zhang, Weifeng Luo, et~al.
\newblock Magi-1: Autoregressive video generation at scale.
\newblock \emph{arXiv preprint arXiv:2505.13211}, 2025.

\bibitem[Theis et~al.(2022)Theis, Salimans, Hoffman, and Mentzer]{theis2022lossy}
Lucas Theis, Tim Salimans, Matthew~D Hoffman, and Fabian Mentzer.
\newblock Lossy compression with gaussian diffusion.
\newblock \emph{arXiv preprint arXiv:2206.08889}, 2022.

\bibitem[Vaisman et~al.(2026)Vaisman, Ohayon, Manor, Elad, and Michaeli]{vaisman2026turboddcm}
Amit Vaisman, Guy Ohayon, Hila Manor, Michael Elad, and Tomer Michaeli.
\newblock Turbo-{DDCM}: Fast and flexible zero-shot diffusion-based image compression.
\newblock In \emph{The Fourteenth International Conference on Learning Representations}, 2026.

\bibitem[Vonderfecht and Liu(2025)]{vonderfecht2025lossy}
Jeremy Vonderfecht and Feng Liu.
\newblock Lossy compression with pretrained diffusion models.
\newblock In \emph{The Thirteenth International Conference on Learning Representations}, 2025.

\bibitem[Wang et~al.(2016)Wang, Gan, Hu, Lin, Jin, Song, Wang, Katsavounidis, Aaron, and Kuo]{wang2016mcl}
Haiqiang Wang, Weihao Gan, Sudeng Hu, Joe~Yuchieh Lin, Lina Jin, Longguang Song, Ping Wang, Ioannis Katsavounidis, Anne Aaron, and C-C~Jay Kuo.
\newblock Mcl-jcv: a jnd-based h. 264/avc video quality assessment dataset.
\newblock In \emph{2016 IEEE international conference on image processing (ICIP)}, pages 1509--1513. IEEE, 2016.

\bibitem[Wang et~al.(2024)Wang, Yue, Zhou, Chan, and Loy]{stablesr}
Jianyi Wang, Zongsheng Yue, Shangchen Zhou, Kelvin~C.K. Chan, and Chen~Change Loy.
\newblock Exploiting diffusion prior for real-world image super-resolution.
\newblock \emph{International Journal of Computer Vision}, 2024.

\bibitem[Wang et~al.(2003)Wang, Simoncelli, and Bovik]{msssim}
Zhou Wang, Eero~P Simoncelli, and Alan~C Bovik.
\newblock Multiscale structural similarity for image quality assessment.
\newblock In \emph{The thrity-seventh asilomar conference on signals, systems \& computers, 2003}, pages 1398--1402. Ieee, 2003.

\bibitem[Wu et~al.(2023)Wu, Zhang, Liao, Chen, Hou, Wang, Sun, Yan, and Lin]{wu2023dover}
Haoning Wu, Erli Zhang, Liang Liao, Chaofeng Chen, Jingwen~Hou Hou, Annan Wang, Wenxiu~Sun Sun, Qiong Yan, and Weisi Lin.
\newblock Exploring video quality assessment on user generated contents from aesthetic and technical perspectives.
\newblock In \emph{International Conference on Computer Vision (ICCV)}, 2023.

\bibitem[Xue et~al.(2025{\natexlab{a}})Xue, Jia, Li, Li, Zhang, and Lu]{xue2025one}
Naifu Xue, Zhaoyang Jia, Jiahao Li, Bin Li, Yuan Zhang, and Yan Lu.
\newblock One-step diffusion-based image compression with semantic distillation.
\newblock In \emph{The Thirty-ninth Annual Conference on Neural Information Processing Systems}, 2025{\natexlab{a}}.

\bibitem[Xue et~al.(2025{\natexlab{b}})Xue, Jia, Li, Li, Zheng, Zhang, and Lu]{xue2025single}
Naifu Xue, Zhaoyang Jia, Jiahao Li, Bin Li, Zihan Zheng, Yuan Zhang, and Yan Lu.
\newblock Single-step diffusion-based video coding with semantic-temporal guidance.
\newblock \emph{arXiv preprint arXiv:2512.07480}, 2025{\natexlab{b}}.

\bibitem[Yang et~al.(2022)Yang, Timofte, and Van~Gool]{yangplvc}
Ren Yang, Radu Timofte, and Luc Van~Gool.
\newblock Perceptual learned video compression with recurrent conditional gan.
\newblock In \emph{IJCAI}, pages 1537--1544, 2022.

\bibitem[Yin et~al.(2025)Yin, Zhang, Zhang, Freeman, Durand, Shechtman, and Huang]{yin2025causvid}
Tianwei Yin, Qiang Zhang, Richard Zhang, William~T Freeman, Fredo Durand, Eli Shechtman, and Xun Huang.
\newblock From slow bidirectional to fast autoregressive video diffusion models.
\newblock In \emph{Proceedings of the IEEE/CVF Conference on Computer Vision and Pattern Recognition}, pages 22963--22974, 2025.

\bibitem[Zeng et~al.(2026)Zeng, Su, Liu, Lu, Tatsumi, and Watanabe]{zeng2026gvcc}
Ziyue Zeng, Xun Su, Haoyuan Liu, Bingyu Lu, Yui Tatsumi, and Hiroshi Watanabe.
\newblock Gvcc: Zero-shot video compression via codebook-driven stochastic rectified flow.
\newblock \emph{arXiv preprint arXiv:2603.26571}, 2026.

\bibitem[Zhang et~al.(2026)Zhang, Li, Chen, Wang, Xu, Li, Sebe, Kwong, and Wang]{zhang2026video}
Pingping Zhang, Jinlong Li, Kecheng Chen, Meng Wang, Long Xu, Haoliang Li, Nicu Sebe, Sam Kwong, and Shiqi Wang.
\newblock When video compression meets multimodal large language models: A unified paradigm for cross-modality video compression.
\newblock \emph{IEEE Signal Processing Letters}, 2026.

\bibitem[Zhang et~al.(2018)Zhang, Isola, Efros, Shechtman, and Wang]{lpips}
Richard Zhang, Phillip Isola, Alexei~A Efros, Eli Shechtman, and Oliver Wang.
\newblock The unreasonable effectiveness of deep features as a perceptual metric.
\newblock In \emph{CVPR}, pages 586--595, 2018.

\bibitem[Zhang et~al.(2025)Zhang, Luo, Li, and Liu]{stablecodec}
Tianyu Zhang, Xin Luo, Li Li, and Dong Liu.
\newblock Stablecodec: Taming one-step diffusion for extreme image compression.
\newblock In \emph{Proceedings of the IEEE/CVF International Conference on Computer Vision (ICCV)}, pages 17379--17389, 2025.

\bibitem[Zhu et~al.(2026)Zhu, Zhao, He, Su, Li, and Zhu]{zhu2026causalforcing}
Hongzhou Zhu, Min Zhao, Guande He, Hang Su, Chongxuan Li, and Jun Zhu.
\newblock Causal forcing: Autoregressive diffusion distillation done right for high-quality real-time interactive video generation.
\newblock \emph{arXiv preprint arXiv:2602.02214}, 2026.

\end{thebibliography}
}

\clearpage
\setcounter{page}{1}
\maketitlesupplementary
\section{Additional Experiments}
\begin{figure*}[t]
    \centering
    \includegraphics[width=\linewidth]{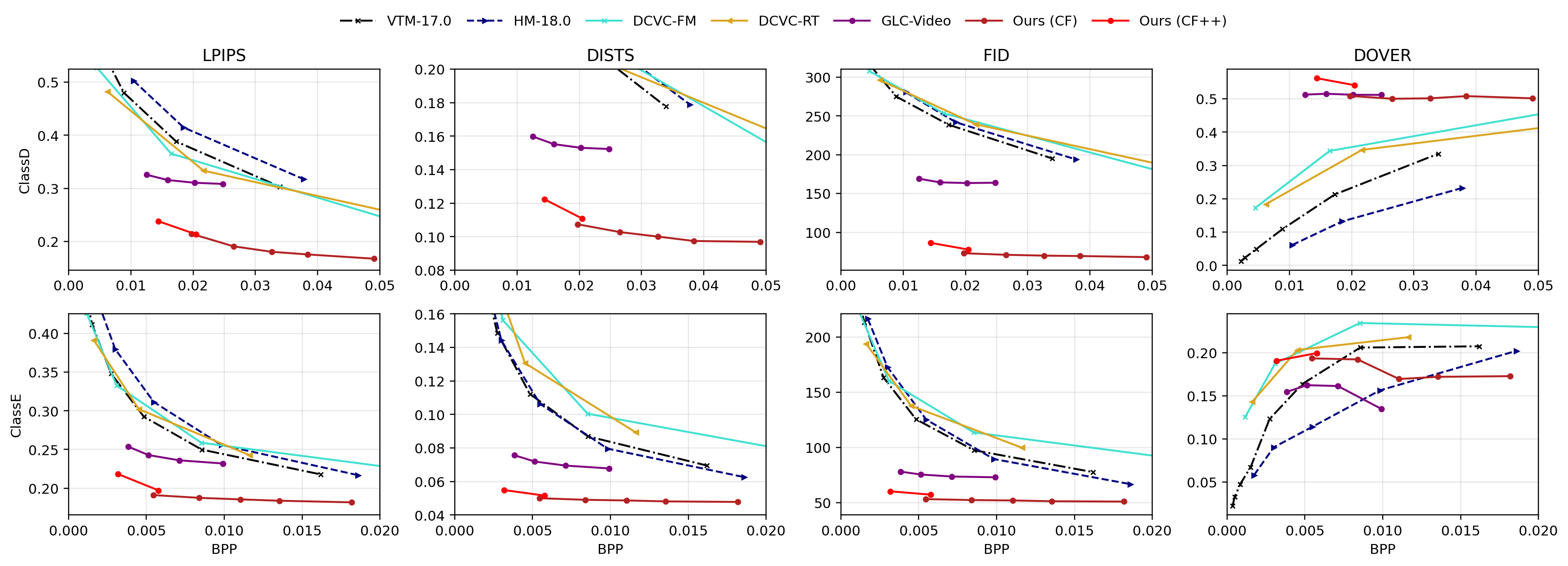}
    \vspace{-2em}
    \caption{Perceptual comparisons on HEVC Class D and Class E.}
    \label{fig:supp_classde_perceptual}
\end{figure*}
\subsection{More Quantitative Results}
Fig.~\ref{fig:supp_classde_perceptual} further presents perceptual results on HEVC Class D and Class E. For the HEVC Class D dataset, whose resolution ($416\times240$) is relatively low, we instead compute FID using $128\times128$ patches. On Class D, ZeroGVC consistently outperforms competing methods across LPIPS, DISTS, FID, and DOVER. Because the original resolution of Class D is too small and does not match the diffusion prior, we resize the videos to 480p for compression, while still computing the bitrate at the original resolution for fair comparison. On Class E, ZeroGVC also achieves clear improvements in LPIPS, DISTS, and FID, while maintaining strong DOVER performance. These results show that ZeroGVC generalizes well to lower-resolution and talking-head video sequences.

\begin{figure*}[t]
    \centering
    \includegraphics[width=\linewidth]{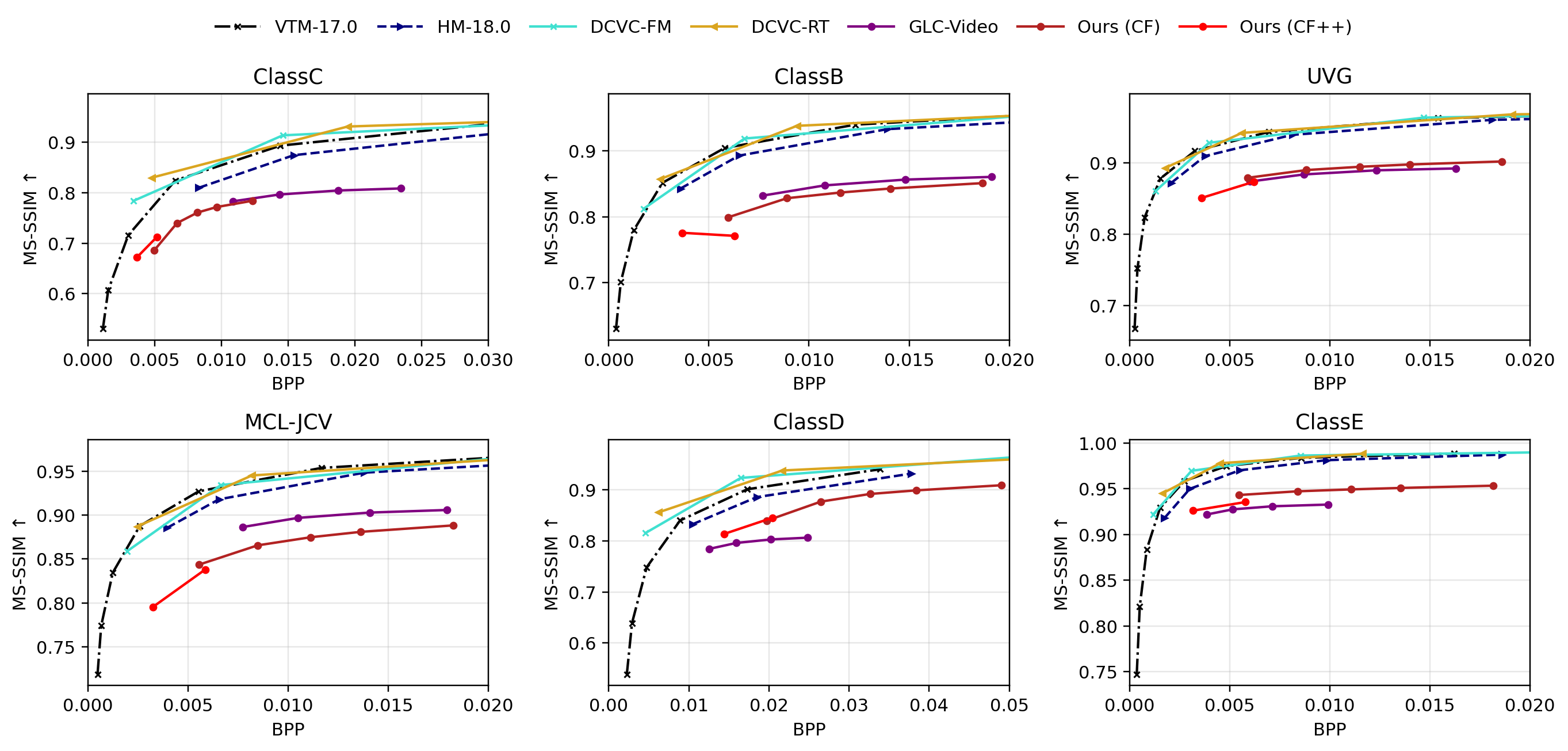}
    \vspace{-2em}
    \caption{MS-SSIM comparisons on all evaluated datasets.}
    \label{fig:supp_msssim}
\end{figure*}

Fig.~\ref{fig:supp_msssim} reports MS-SSIM on all evaluated datasets. ZeroGVC achieves reasonable MS-SSIM and generally remains comparable to or better than the generative baseline GLC-Video.

Table~\ref{tab:supp_perceptual_bd} provides the corresponding BD-Rate and BD-Metric comparisons for perceptual quality. Across HEVC Classes B--E, UVG, and MCL-JCV, ZeroGVC consistently achieves stronger perceptual rate--distortion trade-offs than competing baselines.

\begin{table*}[t]
    \centering
    \caption{BD-Rate (\%) and BD-Metric comparisons on perceptual metrics. Each entry reports BD-Rate / BD-Metric relative to DCVC-RT. Lower BD-Rate values indicate bitrate savings. For LPIPS, DISTS, and FID, lower BD-Metric values are better.}
    \label{tab:supp_perceptual_bd}
    \vspace{-0.5em}
    \scriptsize
    \setlength{\tabcolsep}{2.4pt}
    \renewcommand{\arraystretch}{1.08}
    \resizebox{\textwidth}{!}{
    \begin{tabular}{ll|cc|cc|c|cc}
        \toprule
        Dataset & Metric & HEVC & VVC & DCVC-FM & DCVC-RT & GLC-Video & CF++ & CF \\
        \midrule
        \multirow{3}{*}{Class B}
            & LPIPS$\downarrow$ & 19.58 / 0.0153 & -10.91 / -0.0100 & 6.44 / 0.0049 & 0.00 / 0.0000 & -70.44 / -0.0807 & \tcr{-84.35 / -0.1394} & N/A / -0.1104 \\
            & DISTS$\downarrow$ & -25.34 / -0.0139 & -27.18 / -0.0154 & 7.09 / 0.0032 & 0.00 / 0.0000 & N/A / -0.0717 & \tcr{N/A / -0.1190} & \tcb{N/A / -0.0935} \\
            & FID$\downarrow$ & -40.93 / -24.9695 & -44.03 / -24.0417 & 13.24 / 4.4978 & 0.00 / 0.0000 & N/A / -83.0282 & \tcr{N/A / -116.4824} & \tcb{N/A / -94.4978} \\
        \midrule
        \multirow{3}{*}{Class C}
            & LPIPS$\downarrow$ & 57.18 / 0.0459 & 18.93 / 0.0178 & -1.03 / -0.0012 & 0.00 / 0.0000 & -54.88 / -0.0724 & \tcr{-82.63 / -0.1746} & \tcb{-79.23 / -0.1539} \\
            & DISTS$\downarrow$ & -11.42 / -0.0081 & -12.54 / -0.0089 & 0.79 / 0.0004 & 0.00 / 0.0000 & -78.53 / -0.0714 & \tcr{N/A / -0.1599} & \tcb{N/A / -0.1360} \\
            & FID$\downarrow$ & -35.51 / -23.0681 & -29.35 / -17.8440 & 20.12 / 8.2886 & 0.00 / 0.0000 & N/A / -95.2452 & \tcr{N/A / -156.4355} & \tcb{N/A / -142.8684} \\
        \midrule
        \multirow{3}{*}{Class D}
            & LPIPS$\downarrow$ & 69.22 / 0.0622 & 30.41 / 0.0329 & -0.47 / -0.0007 & 0.00 / 0.0000 & -35.72 / -0.0433 & \tcr{-65.82 / -0.1328} & \tcb{-69.03 / -0.1071} \\
            & DISTS$\downarrow$ & 22.03 / 0.0135 & 3.01 / 0.0019 & -0.03 / -0.0001 & 0.00 / 0.0000 & -65.71 / -0.0683 & \tcr{N/A / -0.1064} & \tcb{N/A / -0.0824} \\
            & FID$\downarrow$ & -8.56 / -5.9537 & -18.60 / -11.7729 & -2.41 / -1.9642 & 0.00 / 0.0000 & -99.33 / -86.5672 & \tcr{N/A / -171.3125} & \tcb{N/A / -144.0222} \\
        \midrule
        \multirow{3}{*}{Class E}
            & LPIPS$\downarrow$ & 36.58 / 0.0309 & -0.07 / -0.0012 & 1.20 / -0.0001 & 0.00 / 0.0000 & -57.95 / -0.0412 & \tcr{N/A / -0.0769} & \tcb{N/A / -0.0766} \\
            & DISTS$\downarrow$ & -16.77 / -0.0134 & -18.83 / -0.0141 & 7.59 / 0.0013 & 0.00 / 0.0000 & N/A / -0.0444 & \tcr{N/A / -0.0763} & \tcb{N/A / -0.0542} \\
            & FID$\downarrow$ & 6.91 / 3.0270 & -6.21 / -4.5391 & 8.58 / 3.4368 & 0.00 / 0.0000 & N/A / -49.5519 & \tcr{N/A / -76.9684} & \tcb{N/A / -61.1942} \\
        \midrule
        \multirow{3}{*}{UVG}
            & LPIPS$\downarrow$ & -2.28 / 0.0005 & -21.98 / -0.0164 & 4.72 / 0.0033 & 0.00 / 0.0000 & -61.29 / -0.0556 & \tcr{-84.46 / -0.0967} & \tcb{-79.22 / -0.0861} \\
            & DISTS$\downarrow$ & -41.93 / -0.0241 & -38.89 / -0.0217 & 3.74 / 0.0022 & 0.00 / 0.0000 & N/A / -0.0857 & \tcr{N/A / -0.1112} & \tcb{N/A / -0.0881} \\
            & FID$\downarrow$ & -38.72 / -19.1270 & -48.41 / -23.8496 & 21.24 / 6.2960 & 0.00 / 0.0000 & N/A / -78.7224 & \tcr{N/A / -96.3871} & \tcb{N/A / -80.2812} \\
        \midrule
        \multirow{3}{*}{MCL-JCV}
            & LPIPS$\downarrow$ & 35.32 / 0.0213 & -11.26 / -0.0085 & 13.15 / 0.0084 & 0.00 / 0.0000 & N/A / -0.0875 & \tcr{-100.00 / -0.1012} & \tcb{N/A / -0.1009} \\
            & DISTS$\downarrow$ & -27.87 / -0.0170 & -36.59 / -0.0232 & 10.01 / 0.0057 & 0.00 / 0.0000 & N/A / -0.1003 & \tcr{N/A / -0.1209} & \tcb{N/A / -0.1045} \\
            & FID$\downarrow$ & -1.66 / -4.8025 & -29.15 / -12.6398 & 25.82 / 6.8514 & 0.00 / 0.0000 & N/A / -84.9948 & \tcr{N/A / -92.2164} & \tcb{N/A / -88.4687} \\
        \bottomrule
    \end{tabular}}
    \vspace{-1em}
\end{table*}
    
\subsection{More Visual Results}
\begin{figure*}[t]
    \centering
    \includegraphics[width=\linewidth]{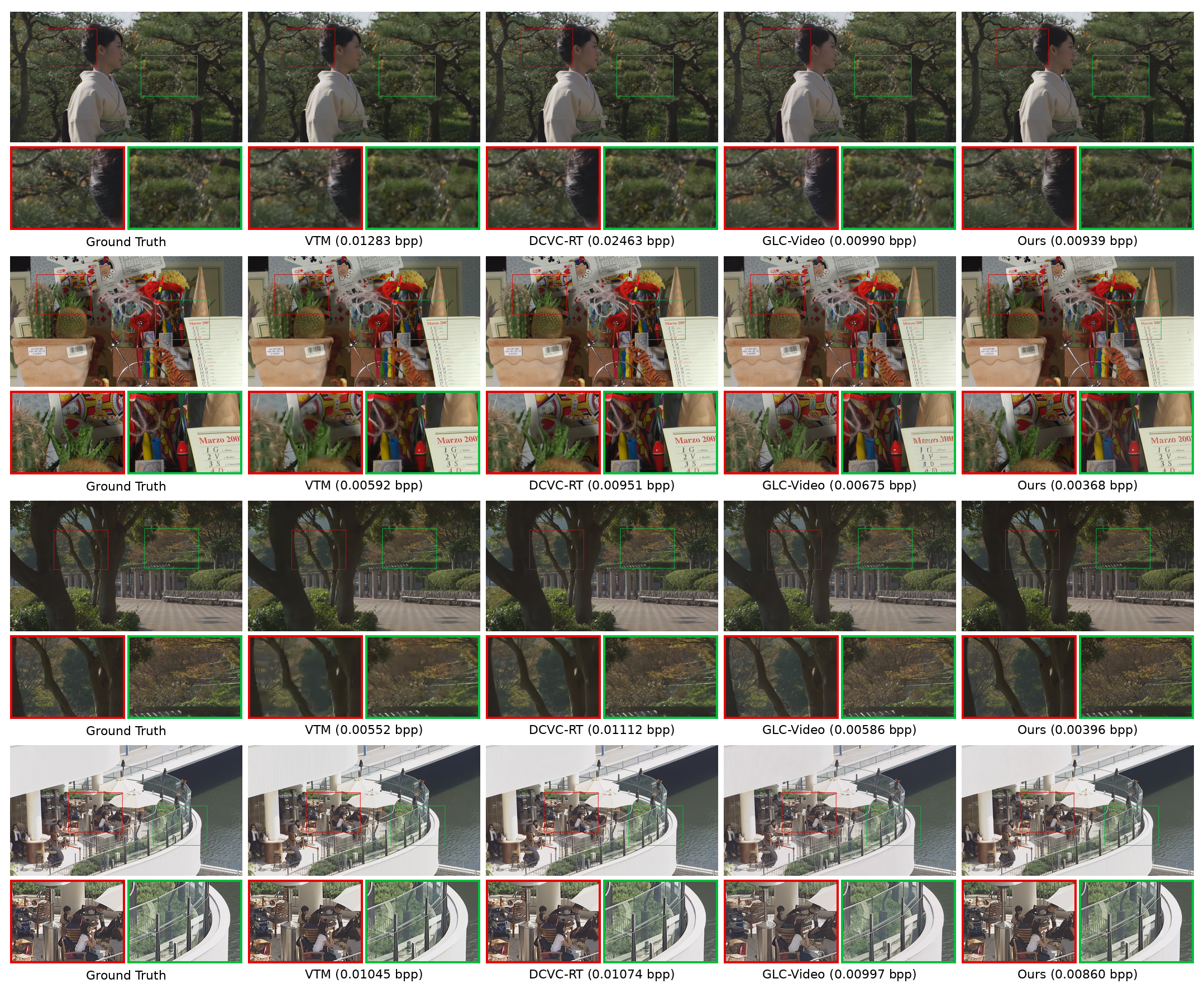}
    \vspace{-1em}
    \caption{More qualitative comparisons with baseline methods.}
    \label{fig:supp_subjective_classb}
\end{figure*}

\begin{figure*}[t]
    \centering
    \includegraphics[width=\linewidth]{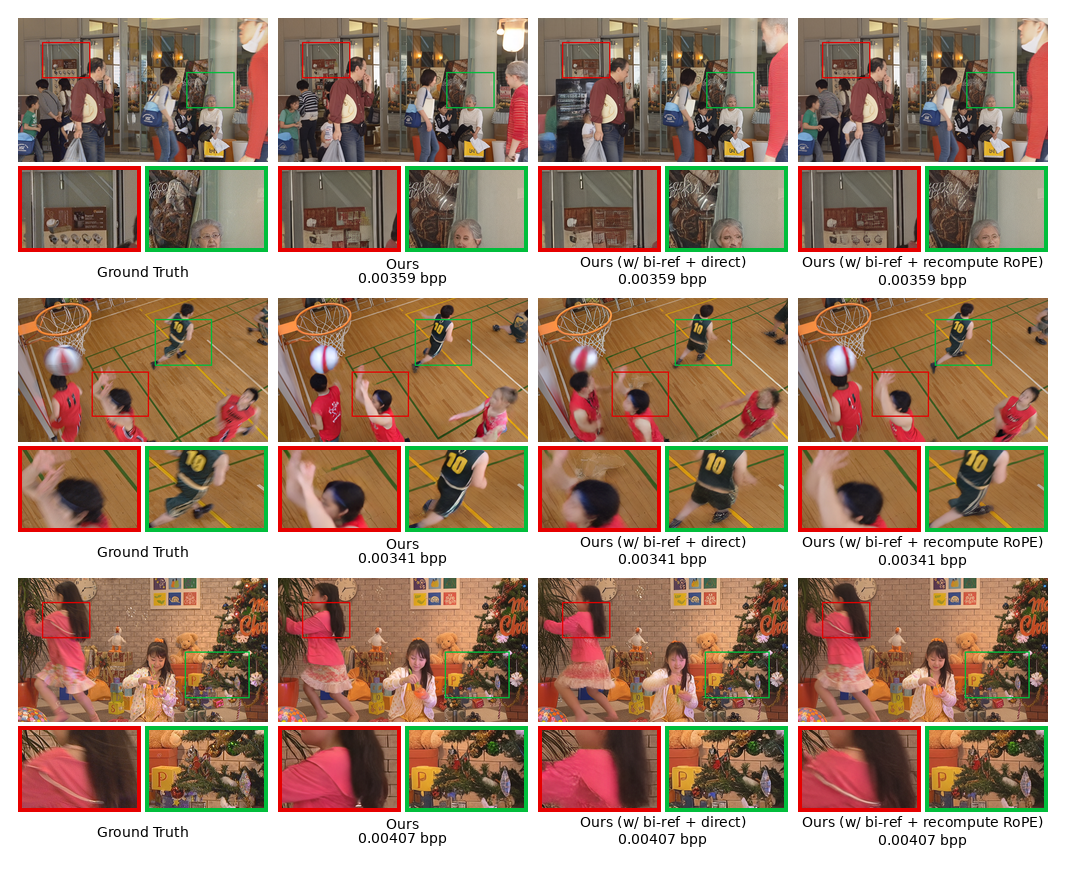}
    \vspace{-1em}
    \caption{More qualitative comparisons of the bidirectional reference mode.}
    \label{fig:supp_bidirectional_subjective}
\end{figure*}

Fig.~\ref{fig:supp_subjective_classb} provides more visual comparisons with baseline methods. Consistent with the observations in the main paper, traditional and neural compression baselines tend to over-smooth fine details or introduce visible artifacts at low bitrates. GLC-Video improves perceptual realism in some regions, but may still lose local structures. In contrast, ZeroGVC better preserves object boundaries, background textures, and fine structures, leading to more realistic reconstructions at comparable or lower bitrates.

Fig.~\ref{fig:supp_bidirectional_subjective} shows additional examples for the bidirectional reference mode. The results follow the same trend as in the main paper: using future reference information helps recover local details that are degraded by autoregressive error propagation. Moreover, recomputing RoPE for the reference tokens produces clearer structures than direct KV cache appending. These visual results further support the effectiveness of the proposed bidirectional reference design.

\end{document}